\documentstyle [11pt]{article} 
\def\pp       {\noindent\parshape 2 0.0 truein 06.5 truein 0.4 truein 06.1
truein}
\def\n {\noindent}

\def\cge   {{$_ >\atop{^\sim}$} }
\def\cle   {{$_ <\atop{^\sim}$} }
\def\Vband {{$V_{606}$} }
\def\Iband {{$I_{814}$} }

\begin{document}

\begin{center}

\Large

\bf{The morphological mix of field galaxies to $m_{I} = 24.25$ mag ($b_{J}\sim
26$ mag) from a deep HST\footnote{Based on observations with the NASA/ESA {\it
Hubble Space Telescope} obtained at the Space Telescope Science Institute,
which is operated by AURA, Inc., under NASA Contract NAS 5-26555.} WFPC2 image}

\vspace{0.5cm}

\large Simon P. Driver, Rogier A. Windhorst, Eric J. Ostrander,

Department of Physics and Astronomy, Arizona State University,

Box 871504, Tempe, AZ\ \ 85287-1504

\vspace{0.5cm}

William C. Keel,

Department of Physics and Astronomy, University of Alabama,

Box 870324, Tuscaloosa, AL\ \ 35487-0324

\vspace{0.5cm}

Richard E. Griffiths and Kavan U. Ratnatunga,

Bloomberg Center for Physics and Astronomy, The Johns Hopkins University,

Box 182695, Baltimore, MD\ \ 21218-2695

\vspace{1.5cm} 

Accepted for publication in:

\vspace{0.5cm}

{\it The Astrophysical Journal (Letters)}, 449, (August 10$^{\mbox{th}}$ 1995)

\vspace{1.0cm}

\end{center}

\normalsize

\pagebreak 

\begin{center} \section*{ABSTRACT} \end{center}

We determine the morphological mix of field galaxies down to $m_{I}\simeq
24.25$
mag ($m_{B}\sim 26.0$ mag) from a single ultradeep HST WFPC2 image in both the
\Vband and \Iband filters. In total, we find 227 objects with $m_{I}\le 24.5$
mag and classify these into three types: ellipticals (16\%), early-type spirals
(37\%) and late-type spirals/Irregulars (47\%). The differential number counts
for each type are compared to simple models in a standard flat cosmology. We
find that both the elliptical and early-type spiral number counts are well
described by {\it little or no}-evolution models, but only when normalized at
$b_{J} = 18.0$ mag. Given the uncertainties in the luminosity function (LF)
normalization, both populations are consistent with a mild evolutionary
scenario
based on a normal/low rate of star-formation. This constrains the end of the
last {\it major} star-formation epoch in the giant galaxy populations to $z\geq
0.8$.

Conversely, the density of the observed late-type/Irregular population is found
to be a factor of 10 in excess of the conventional no-evolution model. This
large population might be explained by either a modified {\it local} dwarf-rich
LF, and/or strong evolution acting on the {\it local} LF. For the dwarf-rich
case, a {\it steep} faint-end Schechter-slope ($\alpha\simeq -1.8$) is required
plus a five-fold increase in the dwarf normalization. For a purely evolving
model based on a {\it flat} Loveday {\it et al.} (1992) LF ($\alpha\simeq
-1.0$), a ubiquitous starburst of $\Delta I\sim$2.0 mag is needed at z$\simeq
0.5$ for the {\it entire} late-type population. We argue for a combination of
these possibilities, and show that for a steep Marzke {\it et al.} (1994) LF
($\alpha\simeq -1.5$), a starburst of $\sim$ 1.3 mag is required at z $\simeq
0.5$ in the entire late-type population, or $\sim 2.0$ mag in $\sim$20\% of the
population.

\vspace{0.25cm}

\n {\it Subject headings:} galaxies: elliptical --- galaxies: spiral ---
galaxies: irregular --- galaxies: luminosity function --- galaxies: evolution

\bigskip

\section{INTRODUCTION}

Over recent years the nature of faint blue field galaxies observed in deep CCD
images has been addressed through two main techniques: galaxy number counts and
faint redshift surveys (Koo \& Kron 1992, KK92). While differential galaxy
counts in short wavebands show an remarkable excess of faint blue galaxies
(Kron 1982; Broadhurst, Ellis \& Shanks 1988, BES), the current faint redshift
surveys show a distribution which is well fit by the standard no-evolution
model (although the LF normalization is incorrect). This conundrum has given
rise to a large number of sophisticated models, most of which introduce an
evolutionary process (see Phillipps \& Driver 1995, for example, and references
therein).

Recent papers from the HST Medium Deep Survey (MDS) have now introduced an
additional constraint: the {\it morphological} classification of faint field
galaxies at HST's resolution of $0.1''$ FWHM. Casertano {\it et al.} (1995,
CRGINOW) give the differential galaxy number counts for HST bulges and disks
separately from the pre-refurbished WF/PC MDS database (13,500 galaxies with
$m_{I_{F785LP}}<21$ mag.
Griffiths {\it et al.} (1994), Driver, Windhorst \& Griffiths (1995, DWG) and
Glazebrook {\it et al.} (1995a, GL95a) use WFPC2 data to extend the
morphological classification of faint field galaxies to $m_{I}\simeq 22.0$ mag.
They find that about one third of the total differential number counts with
$m_{I}\le 22.0$ mag consist of Sd/Irregular galaxies, far more than expected
based on estimates of the local Sd/Irr population. About 35\% of this
population
shows clear signs of star-formation (c.f. DWG). The remaining contribution to
the number counts at $m_{I}\sim 22.0$ mag consists of normal ellipticals and
spirals, which are well fitted by standard mild or no-evolution models (DWG,
GL95a). Here we analyze the 24-orbit WFPC2 exposure of Windhorst \& Keel (1995,
WK), and extend the morphological classification of faint field galaxies to
$m_{I}\simeq 24.25$ mag ($m_{B} \simeq 26$ mag).

\section{WFPC2 DATA REDUCTION AND IMAGE ANALYSIS}

This single deep WFPC2 field has a total of 5.7 hr exposure in both \Vband and
\Iband, and was reduced as described in WK, DWG, and CRGINOW ({\it i.e.}
super-biases, -darks, -sky-flats, and bad-pixel maps from the MDS). Cosmic-ray
removal was achieved by the technique described in Windhorst, Franklin \&
Neuschaefer (1994) updated for WFPC2. Figure 1 (Color Plate 1) shows a color
plate of the entire WFPC2 field. The PC image was not used for the current
study
due to its lower surface brightness (SB) sensitivity. Initial object positions
and total magnitudes were measured using an isophotal object-finding software
of
Ratnatunga {\it et al.} (1995). All objects were examined by eye and recombined
or split as necessary. This yields two object catalogs in $V$ and $I$ which are
complete, based on the turn-over of the counts, to $m_{I}\simeq 25.50$ and
$m_{V}\simeq 26.75$ mag. In total, we find $\sim 600$ and $\sim 850$ objects,
respectively, down to these limits in the three WFC CCD's, implying a surface
density of $4.4\times 10^{5}$ and $6.2\times 10^{5}$ objects deg$^{-2}$. In
this
paper, we consider only those galaxies with $m_{I}\le 24.25$ mag, {\it i.e.}
1.25 mag above our formal completeness limit. Hence, for low SB galaxies, we
become incomplete for objects with isophotal extent \cge 7 arcsec$^{2}$, and
for
luminous bulges at z\cge 1.0. We find no objects down to $m_{I} = 24.25$ mag
larger than 1.0 arcsec$^{2}$ and the current faint redshift surveys --- which
extend to $m_{b_{J}} = 24$ mag --- find very few galaxies with z\cge 1
(Glazebrook {\it et al.} 1995b), although these surveys may also suffer
SB-selection effects. An in-depth discussion of the completeness of HST images
is given by Neuschaefer {\it et al.} (1995).

\section{HST GALAXY NUMBER COUNTS AS FUNCTION OF TYPE }

Figure 2a shows the differential number counts for the HST $I$-band compared to
ground-based data (Tyson 1988; Lilly {\it et al.} 1991; Driver 1994;
Neuschaefer
{\it et al.} 1995). Note the good agreement between the ground-based counts and
our deep HST counts, as well as with previous HST $I$-band counts (DWG and the
WF/PC $I_{F785LP}$-band counts of CRGINOW). The conversion from WFPC2 F814W to
Cousins $I_{C}$ is derived from Holtzman {\it et al.} (1995) and Bahcall {\it
et
al.} (1994) as: \Iband = $ I_{C} + 0.05$ mag, assuming a mean galaxy color of
$(V-I)_{C}\simeq$ 1.5 (Driver 1994) and (\Vband--\Iband)$\simeq 1.0$ mag (DWG).
Our deep WFPC2 $I$-band counts extend the ground-based counts in Fig. 2a with a
best fit slope of $0.34\pm 0.03$ for $22.0<m_{I}<25.5$ mag. The $V$-band counts
have a slope of $0.38\pm 0.03$ for $23.0<m_{V}<27.0$ mag, implying a trend
towards bluer $(V-I)$ colors at fainter magnitudes. At $m_{I}\simeq 21.5$ mag,
the mean field galaxy $(V-I)$ color is $1.3\pm 0.3\ (1\sigma)$, changing to
$0.8\pm 0.3$ at $m_{I}\simeq 25.0$ mag, and shows little correlation with
galaxy
type (see DWG). Hence, we do not use color to classify, moreover redshifts and
K-corrections at I$\simeq$24 mag ($b_{J}\simeq$26) are unknown.

Following the methods in DWG, we did the photometry for the 227 objects with
{\it total} flux $m_{I}\leq 24.5$ mag, resulting in an accuracy of $\pm0.1$
mags
(c.f. DWG). The objects were classified into ellipticals, early-type spirals,
and late-type spirals/Irregulars using both the $V$ and $I$-band morphology and
light-profiles. As in DWG the distinction  between ``early'' (Sabc) and
``late''
(Sd/Irr) is based on a combination of their measured central SB and a careful
assessment of their measured light-profiles, bulge-to-disk ratios, and
grey-scale images (see DWG for an example of the Hubble sequence at
$m_{I}\simeq
21.5$ mag). The final assigned galaxy types are the consensus of four {\it
independent} classifiers (SPD, RAW, EJO and WCK), and were determined by
averaging the number of galaxies in each class and in each magnitude interval
(as opposed to defining an average or most likely class for each object
individually). The consistency between the four independent classifiers is
summarized in Table 1. Note that the scatter in the interval
$22.25<m_{I}<24.25$ is equivalent to that in $20.25<m_{I}<22.25$mag.

Figure 2 shows the differential number counts for: ellipticals (2b), early-type
spirals (2c), and late-type spirals/Irregulars (2d). Fig. 2 also includes the
WF/PC data of CRGINOW (for all galaxies, ellipticals and spirals) for $I\leq
21.0$ mag, and the WFPC2 data of DWG for $I\leq 22.0$ mag. Errors are derived
from assuming Poisson statistics. The formal errors ($\sqrt{n}$) are shown by
{\it solid} errorbars. Vertical dotted lines indicate the {\it total} range
covered by the four independent classifiers, and are a {\it conservative}
estimate of the true classification errors. Where our new data overlaps with
that of CRGINOW and DWG, there is good agreement within the formal errors. The
largest errors occur for the E/S0's, where the statistics are smallest and
which are the hardest to recognize --- even with HST --- due to their small
scale-lengths (Mutz {\it et al.} 1994; Windhorst {\it et al.} 1994a, b). The
level of agreement between our four independent classifiers down to I$\le$
24.25
mag is comparable to that achieved by DWG for $I\le$ 22.0 mag. In Table 1, the
disagreement between classifiers is exaggerated because the data is binned into
a {\it small} number of classes. E.g., if two classifiers agree on a galaxy as
S0 and two as Sa, this might be considered a reasonable agreement, but because
we bin objects into three classes, such disagreements seem more significant.
Table 2 shows the morphological mixes observed at various flux limits. The
bright end of our HST sample agrees well with the faint end of the DWG sample,
and the bright end of the DWG sample agrees well with the local fractions
(Shanks {\it et al.} 1984).

We applied the bulge {\it or} disk-fitting algorithms of CRGINOW (1995) to
provide an {\it independent} automated classification for all 227 objects into
ellipticals or spirals, as shown in Table 2. While this algorithm cannot
distinguish between mid and late-type spirals, the fraction of brighter
galaxies
($21\le I<23.0$ mag) classified into bulges and disks {\it by the algorithm} is
30\% and 70\%, respectively, while {\it by eye} it is 32\% and 68\%. At the
faint end ($23\le I<24.5$ mag) the correspondence is still satisfactory: the
bulge and disk fractions are 21\% and 79\% {\it by the algorithm}, and 14\% and
86\% {\it by eye}, respectively. Software to achieve two-dimensional
simultaneous bulge {\it plus} disk fits is under development, and will be
described in a subsequent paper (Ratnatunga {\it et al.} 1995). A comprehensive
spectroscopic survey is also in progress for the DWG sample (Driver
{\it et al.} 1995, in prep.), and will help confirm these classifications.

We conclude that we have --- within the limits of the available data --- a
consistent picture of the trend of morphological mix with apparent magnitude.
Figure 2 and Table 2 show that the galaxy mix at {\it faint} magnitudes is
significantly different from the local values, with a far higher proportion of
late-type galaxies is seen at the HST limit.

\section{MODEL FITTING}

To model the differential number counts for each galaxy type separately, we
adopted three independent luminosity functions (LF's), a standard flat
cosmology
($\Lambda=0, \Omega=1$ and $H_{o}=50$ kms$^{-1} Mpc^{-1}$, c.f. Phillipps,
Davies \& Disney 1990) and K-corrections (c.f. Driver {\it et al.} 1994). The
modelling process is described in DWG. For {\it non-evolving} E/SO's and
Sabc's,
we parameterized the {\it observed} luminosity distributions (LD's) from Marzke
{\it et al.} (1994b, MGHC) using their tabulated values for E/SO: $M_{B}^{*} =
-20.5$ mag, $\alpha = -0.9$, $\phi_{*} = 1.14\times 10^{-3}\ Mpc^{-3}$, and for
Sabc: $M_{B}^{*} = -20.3$ mag, $\alpha = -0.8$,
$\phi_{*} = 1.74\times 10^{-3}\ Mpc^{-3}$.
The LF's were exponentially cut off at $M_{B}^{cut}$ = $-17.5$ mag
[{\it i.e.} the Schechter (1976) function was multiplied by
$exp(-10^{0.4(M-M^{cut})})$]. To convert the observed $B$-band LD's to the
$I$-band, we adopted the following mean colors (Windhorst {\it et al.} 1994b)
at z=0: $(B-I)_{E/S0}\simeq 2.3$, $(B-I)_{Sabc}\simeq 1.9$, and
$(B-I)_{Sd/Irr}\simeq 1.4$ mag. The LF models were normalized to the observed
number counts for all types at $m_{b_{J}}\simeq 18-20$ mag (c.f. KK92),
resulting in a uniform 0.3 dex increase in numbers for each type. As discussed
in DWG, this discrepancy in normalization arises from the observed steep galaxy
counts at $b_{J}$\cle 17 mag (see also Shanks {\it et al.} 1989). In line with
standard practise (c.f., KK92), we normalize our models, intended to represent
a sufficient distance to cover a homogeneous volume (i.e., $z\sim 0.15$ or
$b_{J}\sim 18-20$ mag), but at a low enough redshift that strong evolution has
not yet occurred.

Figure 2b and 2c show the model predictions for ellipticals and early-type
spirals, compared to our HST data. The models mimic the observed distribution
rather well, implying --- with the adopted LF normalization --- little
luminosity evolution (LE) for both the elliptical and early-type spiral
populations. Note that the lightly dotted line on Figures 2b and 2c represent
the predictions from the {\it un}normalized local LF's, which would imply
stronger evolution for the early type populations. To model the
late-type/Irregular population, three alternate LF's were used in the following
four models:

\n {\bf (a)} a {\it no-evolution} prediction based on the Loveday {\it et al.}
(1992, LPEM) LD for late-type galaxies ($M_{B}^{*} = -18.5$ mag, $\alpha =
-1.1$, $\phi_{*}(Sd/Irr) = 7.0\times 10^{-4}$Mpc$^{-3}$, c.f. DWG).

\n {\bf (b)} a {\it no-evolution} prediction based on the MGHC-LD for late-type
galaxies ($M_{B}^{*} = -20.3$ mag, $\alpha = -1.5$, $\phi_{*} = 2.5 \times
10^{-4}$, DWG).

\n {\bf (c)} an {\it evolving} model (Phillipps \& Driver 1995) that mimics a
ubiquitous starburst in the Sd/Irr galaxy population at z $\simeq 0.5$, after
which their luminosity declines exponentially with time ({\it i.e.} $\Delta m
\propto\beta [1 - (1+z)^{-\frac{3}{2}}]$ for $\Omega = 1$ with $M_{B}^{*}$,
$\alpha$, and $\phi_{*}(Sd/Irr)$ from (a) or (b), and $\beta=$free).

\n {\bf (d)} a {\it no-evolution} dwarf-rich model with $\alpha = -1.8$, (c.f.
Driver {\it et al.} 1994). The normalization is {\it increased} until a fit is
found to the data ($M_{B}^{*} = -18.0$ mag, $\alpha = -1.8$, and
$\phi_{*}(Sd/Irr)=$ free).

For models (a) and (b), the late-type/Irregular parameters listed above were
taken {\it directly} from DWG. Figure 2d shows that they grossly under-predict
the observed late-type/Irregular population. For model (c), the amount of
luminosity evolution (LE) required for the {\it entire} late-type population
was
found to be $\Delta m\sim 2.0$ mag at z $\simeq 0.5$ to match the counts with
the LPEM-LF, and $\sim 1.3$ mag with the MGHC-LF. The required $\sim 1.3$ mag
increase in luminosity for the {\it entire} late-type galaxy population is
equivalent to a $\sim2.0$ mag increase in $\sim 20$\% of the population. For
the
dwarf-rich model (d), a local normalization of $\phi_{*}\sim 3.5\times 10^{-2}\
Mpc^{-3}$ was required to match the counts, which is a factor of 5 greater than
that of LPEM (not including the additional 0.3 dex normalization). Such a
dwarf-LF, however, is inconsistent with the faint redshift surveys, as it
predicts too many low-redshift objects (Driver 1994; Phillipps \& Driver 1995).

Figure 3a (Color Plate 2) compares the best-fit individual LF's adopted for
each
galaxy type. Figure 3b shows the conventional differential number counts with
the contribution from each galaxy type indicated by the best-fit model lines
from Figs. 2a--2d. Fig. 3c shows the observed LD, which reflect the {\it
actual}
relative numbers that would be observed in a {\it magnitude-limited} survey
(assuming equal selection effects for all types). Fig. 3d shows the {\it
normalized} differential galaxy number counts which emphasizes the differences
between the models and the data.

\section{DISCUSSION AND CONCLUSIONS}

Our observed differential HST counts for ellipticals (E/S0's) and early-type
spirals (Sabc's) agree with the simple {\it no-evolution} model shown here,
assuming the LF normalization is correct. Our faintest HST counts lie at best
marginally above the model predictions. This is consistent with a mildly
evolving ``giant'' galaxy population undergoing a normal rate of star-formation
(e.g. Bruzual \& Charlot 1993).
As a caveat to the above we note that the models described here have been
normalized at $b_{J} = 18.0$ mag. Without this normalization, evolution {\it
is}
required in the luminous populations to explain the $\sim$0.3 dex difference
between the models and our data. The fact that both populations require the
same
normalization and that the shape of the observed distribution matches the shape
of the models rather well perhaps argues for the case to normalize. {\it
However, until the local predictions of faint galaxy models are reconciled with
the local redshift surveys, moderate (local) evolution in the giant galaxy
populations cannot be ruled out.} Even so, an agreement within 0.3 dex places
the end of the {\it major} star-formation epoch for these types to $z$\cge 0.8,
which is consistent with the lack of scale-length evolution observed in HST
ellipticals out to z=0.8 (Mutz {\it et al.} 1994).

The late-type/Irregular population shows a considerable discrepancy between the
no-evolution predictions and our deep HST data. This was also noted by DWG and
GL95a down to I$\le$22.0 mag, and has now been confirmed down to $m_{I}\sim
24.25$ mag with a steeply rising --- almost Euclidean --- slope and {\it no}
indication for a turnover. The {\it no-evolution} models based on either the
LPEM or MGHC LF's fall short of our deep HST observations by up to a factor 10
at the faintest limits. This late-type population is therefore clearly
responsible for the ``faint blue galaxy excess''. Two possible solutions to
this
discrepancy are: strong evolution and/or a serious underestimation of the local
space density of dwarf galaxies. The possibility of large-scale non-homogeneity
or a gross error in the cosmological model is ruled out, as the E/S0 and Sabc
models fit reasonably well to our HST data given our basic assumptions. While
either of the possibilities considered here can be forced to fit our data, the
implications are somewhat unpalatable.\footnote{Note that again {\it if}
normalization of the faint galaxy models at $b_{J} = 18$ mag is ignored the
problem is amplified by another 0.3 dex.} If we evolve the widely adopted
LPEM-LF, then $\sim$2.0 mag of brightening would be required at $z\sim 0.5$ in
the {\it entire} dwarf galaxy population! Alternatively, the additional dwarfs
required to explain this population without any evolution result in a
significant low-redshift excess in the faint galaxy redshift distributions,
which is not observed (GL95a), although the statistics in the redshift surveys
are still small and the selection effects formidable. The more recent MGHC-LF
--- based on Zwicky magnitudes --- offers a compromise: if the observed steep
faint-end LF slope undergoes luminosity evolution, a more reasonable value of
$\sim 1.3$ mag brightening is implied at $z\sim 0.5$, equivalent to $\sim 2.0$
mag in $\sim$20\% of the population. The WFPC2 morphological surveys of DWG and
GL95a concur that $\sim$40\% of the late-type/Irregular population shows
evidence for recent star-formation, while the remaining galaxies appear inert.
Together with our deeper HST counts, this leads us to conclude that the
observed
faint blue galaxy excess is caused by a combination of {\it strong evolution}
in a {\it substantial fraction} of the late-type galaxy population {\it coupled
with} an {\it under-representation} of late-type dwarf galaxies in local
surveys (c.f. Driver \& Phillipps 1995). This ``family'' of faint galaxy models
is explored in detail in Phillipps \& Driver (1995).

\vspace{0.2cm}

We thank Doug VanOrsow and Dan Golombek for help in obtaining the data, and
Paul Scowen and Barbara Franklin in creating the color image. We acknowledge
support from HST grants GO.5308.01.93A (RAW),\ .02.93A (WCK), and
GO.2684.03.93A (SPD, RAW, EJO),\ .01.93A (REG, KUR) and EPSCoR grant
EHR-9108761 (WCK).

\bigskip

\section*{REFERENCES}

\baselineskip=15pt 

\begin{description}

\item Bahcall, J. N., {\it et al.} 1994, ApJL, {\bf 435}, L51

\item Broadhurst, T. J., Ellis, R. S., \& Shanks, T.
1988, MNRAS, {\bf 235}, 827 (BES)

\item Broadhurst, T. J., {\it et al.} 
1992, Nature, {\bf 355}, 827

\item Bruzual A., G., \& Charlot, S. 1993, ApJ, {\bf 405}, 538

\item Casertano, S., Ratnatunga, K. U., Griffiths, R. E., Im, M., Neuschaefer,
L. W., Ostrander, E. J., \& Windhorst, R. A. 1995, ApJ, 453, 599 (CRGINOW)

\item Driver, S. P. 1994, PhD Thesis, University of Wales

\item Driver, S. P., {\it et al.} 1994, MNRAS, {\bf 266}, 155

\item Driver, S. P., \& Phillipps, S. 1995, ApJ, submitted

\item Driver, S. P., Windhorst, R. A., \& Griffiths R. E. 1995, ApJ, 453, 48
(DWG)

\item Griffiths, R. E., {\it et al.} 1994, ApJL, {\bf 435}, L019

\item Glazebrook, K., {\it et al.} 1995a, Nature, submitted (GL95a)

\item Glazebrook, K., {\it et al.} 1995b, MNRAS, {\bf 273}, 157

\item Holtzman, J. A. {\it et al.} 1995, PASP, {\bf 107}, 156

\item Im, M., Casertano, S., Griffiths, R. E., Ratnatunga, K.
1995, ApJ, {\bf 441}, 494

\item Koo, D. C., \& Kron, R. G. 1992, ARA\&A, {\bf 30}, 613 (KK92)

\item Kron R. G. 1982, Vistas in Astronomy, {\bf 26}, 37

\item Lilly, S. J., Cowie, L. L., \& Gardner, J. P. 1991, ApJ, {\bf 369}, 79

\item Loveday, J., Peterson, B. A., Efstathiou, G., \& Maddox, S. J. 1992, ApJ,
{\bf 390}, 338 (LPEM)

\item Marzke, R. O., Huchra, J. P., \& Geller, M. J. 1994a, ApJ, {\bf 428}, 43

\item Marzke, R. O., Geller, M. J., Huchra, J. P., \& Corwin Jr, H. G. 1994b,
AJ, {\bf 108}, 437 (MGHC)

\item Mutz, S. B., {\it et al.} 1994, ApJL, {\bf 434}, L55

\item Neuschaefer, L. W., \& Windhorst, R. A. 1995, ApJS, {\bf 96}, 371 (NW95)

\item Neuschaefer, L. W., {\it et al.} 1995, PASP, in press.

\item Phillipps, S., \& Driver, S. P. 1995, MNRAS, 274, 832

\item Phillipps, S., Davies, J. I., \& Disney, M. J. 1990, MNRAS, {\bf 242},
235

\item Ratnatunga, K. U., {\it et al.} 1995, ApJ, in prep.

\item Schechter, P. 1976, ApJ, {\bf 203}, 297

\item Shanks, T., {\it et al.} 1984, MNRAS, {\bf 206}, 767

\item Shanks, T., {\it et al.} 1989, in {\it The Extragalactic Background
Light}, eds Bowyer, S. C. \& Leinert, C., Kluwer Academic Press, p 269

\item Tyson, J. A. 1988, A J, {\bf 96}, 1

\item Windhorst, R. A., {\it et al.} 1994a, AJ, {\bf 107}, 930

\item Windhorst, R. A., {\it et al.} 1994b, ApJ, {\bf 435}, 577

\item Windhorst, R. A., Franklin, B. E., \& Neuschaefer, L. W.
1994c, PASP, {\bf 106}, 798

\item Windhorst, R. A., \& Keel, W. C. 1995, ApJL, submitted

\end{description}

\pagebreak

\vspace{0.5cm} 

\section*{TABLES} 

\pp Table 1: Agreement between the four independent eyeball classifiers as
function of flux.

\smallskip

\begin{tabular}{l|c|c|c}
\hline
\hline
\n Agree  & \multicolumn{3}{c}{Magnitude Interval} \\ \cline{2-4}
\n -ment  &$I<20.25$&$20.25<I<22.25$&$22.25<I<24.25$\\
\hline
\smallskip
\n 4 of 4 & 9 (75\%)& 12 (36\%)     & 53 (36\%) \\
\n 3 of 4 & 3 (25\%)& 12 (36\%)     & 54 (37\%) \\
\n 2 of 4 & 0 ( 0\%)&  9 (28\%)     & 39 (27\%) \\
\hline
\end{tabular}

\vspace{0.5cm}

\pp Table 2: Comparison of the morphological mix of field galaxies by various
groups.

\smallskip

\begin{tabular}{l|c|r|r|r|r}
\hline
\hline
\n Source                        & magnitude        & E/S0 & Sabc & Sd/Irr
& Uncl. \\
\hline
\smallskip
\n MGHC (CFA1)$^{1}$             & $m_{Z}<14.5$     & 35\% & 54\% & 10\%
& 1\% \\
\n MGHC (CFA2)$^{1}$             & $m_{Z}<15.5$     & 42\% & 48\% & 8\%
& 2\% \\
\n Shanks {\it et al.} (1984)    & $m_{b_{J}}<16.0$ & 43\% & 45\% & 12\%
& 0\% \\
\n Griffiths {\it et al.} (1994) & $m_{I}<22.25$    & 19\% & 44\% & 13\%
& 25$^{2}$\% \\
\n DWG bright                    & $m_{I}\sim 20.25$& 36\% & 50\% & 14\%
& 0\% \\
\n GL95a bright                  & $m_{I} = 20.25$  & 31\% & 48\% & 21\%
& 0\% \\
\n DWG faint                     & $m_{I}\sim 21.75$& 28\% & 35\% & 31\%
& 6\% \\
\n GL95a faint                   & $m_{I} = 21.75$  & 29\% & 35\% & 33\%
& 3\% \\
\n Current, bright, Eye          & $m_{I}\sim 22.00$& 32\% & 38\% & 30\%
& 0\% \\
\n Current, bright, Auto$^{3}$   & $m_{I}\sim 22.00$& 30\% &
\multicolumn{2}{c|}{70\%}& 0\% \\
& 70\% & 0\% \\
\n Current, faint, Eye           & $m_{I}\sim 24.00$& 14\% & 30\% & 56\%
& 0\% \\
\n Current, faint, Auto$^{3}$    & $m_{I}\sim 24.00$& 21\% &
\multicolumn{2}{c|}{79\%}& 0\% \\
$\longleftarrow$& 79\% & 0\% \\
\multicolumn{2}{c|}{75\%}& 0\% \\
\hline
\end{tabular}

\smallskip

\n {\it Notes:} $^{1}$ The mix for the CFA1 and CFA2 samples were estimated
from Figure 1 of MGHC.

\n $^{2}$ Griffiths {\it et al.} (1994) note that a significant fraction of the
unclassified galaxies were classified as S0 by one of their two classifiers.

\n $^{3}$ Automated classifications were determined from the profile fits of
CRGINOW and Ratnatunga {\it et al.} (1995). This algorithm cannot distinguish
between Sabc's and Sd/Irr's, so {\it all} disk dominated galaxies are listed
together.

\pagebreak 

\baselineskip=18pt 

\section*{FIGURE CAPTIONS}

\bigskip

\n {\bf Figure 1, COLOR PLATE 1} --- This color image shows the 5.7 hr WFPC2
images in both the \Iband and the \Vband filters surrounding the weak radio
galaxy 53W002 at z=2.390 (Windhorst \& Keel 1995). Images in $V$, $(V+I)/2$ and
$I$ are shown in the blue, green and red guns respectively. A total of 600
objects in the $I$ and 850 objects in the $V$ frames have been detected down to
$m_{I} \simeq 25.5$ and $m_{V}\simeq 26.75$ mag in 0.00136 deg$^{2}$, implying
a surface density of $4.4\times10^{5}$ and $6.2\times10^{5}$ deg$^{-2}$ in $I$
and $V$, respectively. The $V$ and $I$ images have a 6.73$^{o}$ difference in
HST roll angle, leading to apparently extreme object colors at the CCD-edges.

\bigskip

\n {\bf Figure 2} --- The differential $I$-band number counts for: (a) all
galaxies; (b) ellipticals; (c) early-type spirals; and (d) late-type
spirals/Irregulars. The open symbols without errorbars are from the larger but
shallower HST surveys of CRGINOW (small symbols) and DWG (large symbols). The
solid symbols with errorbars represent our new HST WFPC2 data. The errors are
$\sqrt{n/N}$, where $n$ represents number of objects and N the number of
classifiers. The vertical {\it dotted} lines represent the range of agreement
between classifiers, and is a conservative upper limit to the true errors. The
lightly dotted lines on Fig. 2 b and c represent the {\it un}-normalized LF
predictions (see text).

\bigskip

\n {\bf Figure 3, COLOR PLATE 2} --- {\bf (a)} The $I$-band luminosity
functions
(LF's) used in the morphological modelling. Three variants are shown for the
late-type spiral/Irregular population: Loveday {\it et al.} (1992); Marzke {\it
et al.} (1994b), and Driver {\it et al.} (1994a). {\bf (b)} The resulting
differential I-band number counts, along with the best fit models for each type
(blue solid = dwarf-rich case, blue dashed = evolving Marzke-LF), as derived
from Fig. 2a--2d. {\bf (c)} As Fig. 3a, except that the LF's are displayed as
they would be observed in a {\it magnitude-limited} sample. {\bf (d)} The
combined {\it normalized} differential galaxy counts for {\it all} galaxy
types,
and for ellipticals, early and late-type spirals based on our HST images. Also
shown are the ``shallower'' data of CRGINOW and DWG, as well as the best fit
models of Fig. 2a--2d. Note the similarity in shape between the luminosity
distributions of the three major galaxy types in Fig. 3c and their {\it
normalized} differential galaxy counts in Fig. 3d.

\end{document}